\documentclass[12pt]{article}
\usepackage{epsfig}

\def\KK{$\bar{K}^0$-$K^0$~}
\def\BB{$\bar{B}^0$-$B^0$~}
\def\BBs{$\bar{B}_s^0$-$B_s^0$~}

\def\simlt{\stackrel{<}{{}_\sim}}
\def\simgt{\stackrel{>}{{}_\sim}}

\def\bbuildrel#1_#2^#3{\mathrel{\mathop{\kern 0pt#1}\limits_{#2}^{#3}}}
\def\slash#1{\setbox0=\hbox{$#1$}#1\hskip-\wd0\dimen0=5pt\advance
       \dimen0 by-\ht0\advance\dimen0 by\dp0\lower0.5\dimen0\hbox
         to\wd0{\hss\sl/\/\hss}}

\def\be{\begin{eqnarray}}
\def\ee{\end{eqnarray}}

\evensidemargin=\oddsidemargin

\begin{document}

\begin{flushright}
{hep-ph/0108226}
\end{flushright}

\vskip 3mm

\begin{center}
  
  {\bf \boldmath{$\Delta$$F$$=$$2$} processes in the MSSM in large
    \boldmath{$\tan\beta$} limit\footnote{to appear in the proceedings
      of the 9th International Conference on Supersymmetry and
      Unification of Fundamental Interactions SUSY'01, June 11-17,
      Dubna, Russia.}}

  \vskip 5mm
  
  JANUSZ ROSIEK\footnote{On leave of absence from the Institute of
    Theoretical Physics, Warsaw University.}
  
  \vskip 2mm
  
  {\sl Physik Department, Technische Universit{\"a}t M{\"u}nchen,
    D-85748 Garching, GERMANY}\\

\end{center}

\vskip 1mm

\begin{abstract}{We discuss corrections to $\Delta M_{B_d}$, $\Delta
  M_{B_s}$ and to the CP violation parameter $\varepsilon$ in two
  examples of (generalized) minimal flavour violation models: 2HDM and
  MSSM in the large $\tan\beta$ limit.  We show that for $H^+$ not too
  heavy, $\Delta M_{B_s}$ in the MSSM with heavy sparticles can be
  substantially smaller than in the SM due the charged Higgs box
  contributions and in particular due to the growing like
  $\tan^4\beta$ contribution of the double penguin diagrams involving
  neutral Higgs boson exchanges.}
\end{abstract}

\vskip 4mm

The determination of the CKM parameters is the hot topic in particle
physics.  In view of forthcoming precise results from B-factories, it
is particularly important to discuss possible effects of new physics
contributions to $\Delta F=2$ processes: $\Delta M_{B_d}$ and $\Delta
M_{B_s}$ mass differences and to the \KK CP violation parameter
$\varepsilon$.  In general models of new physics fall into the two
following broad classes:\\[1mm]
{$\bullet$} Models in which the CKM matrix remains the unique source
of flavour and CP violation - so-called minimal flavour violation
(MFV) models~\cite{CIDEGAGI,BUGAGOJASI} and their generalizations
(GMFV models).  In the GMFV models the non SM-like operators
contribute significantly to the effective low energy 
Hamiltonian~\cite{BUCHROSL}.\\[1mm]
{$\bullet$} Models with entirely new sources of flavour and/or CP
violation.\\[1mm]
On the basis of the analysis~\cite{BUCHROSL} we discuss here two
examples of the GMFV models: the large $\tan\beta$ limit of the
2HDM(II) and of the MSSM, in which the CKM matrix is the only source
of flavour and CP violation (see e.g.~\cite{MIPORO}).

The effective weak Hamiltonian for $\Delta F=2$ transitions in the
GMVF models can be written as follows
\be\label{heff}
H_{\rm eff}^{\Delta {\rm F}=2} = {G_F^2M_W^2\over16\pi^2}
\sum_i V^i_{\rm CKM} C_i(\mu) Q_i~.
\ee
where $Q_i$ are the set of 8 dimension six $\Delta F=2$
operators~\cite{BUJAUR} and $V^i_{\rm CKM}$ are the appropriate CKM
factors.  $\Delta M_{B_d}$,$\Delta M_{B_s}$ and $\varepsilon$ can be
expressed in terms of, respectively, three real functions~\cite{ALLO}
\be
F^d_{tt}=S_0(x_t) \lbrack 1+f_d\rbrack~,
\qquad
F^s_{tt}=S_0(x_t) \lbrack 1+f_s\rbrack~,
\qquad
F^\varepsilon_{tt}=S_0(x_t) \lbrack 1+f_\varepsilon\rbrack~,
\ee
$S_0(x_t\equiv \bar m_t^2/M_W^2)\approx2.38\pm0.11$ for $\bar
m_t(m_t)=(166\pm5)$ GeV.  We have then
\begin{equation}
\Delta M_q = {G_{\rm F}^2M_W^2\over6\pi^2}M_{B_q} \eta_B 
\hat B_{B_q} F_{B_q}^2  |V_{tq}|^2 F_{tt}^q,
\qquad q=d,s
\label{eqn:xds}
\end{equation}
The functions $F_{tt}^d$, $F_{tt}^s$ and $F_{tt}^\varepsilon$ can be
expressed in terms of $C_i$ as
\begin{eqnarray}\label{hds2}
F_{tt} &=& \left[S_0(x_t)+ \frac{1}{4 r}C_{\rm new}^{\rm VLL}(\mu_t)\right]
+{1\over4r}C^{\rm VRR}_1(\mu_t)+
\bar P_1^{\rm LR} C^{\rm LR}_1(\mu_t) 
+\bar P_2^{\rm LR} C^{\rm LR}_2(\mu_t) \nonumber\\
&&+\bar P_1^{\rm SLL}\left[C^{\rm SLL}_1(\mu_t)+C^{\rm SRR}_1(\mu_t)\right]
+\bar P_2^{\rm SLL}\left[C^{\rm SLL}_2(\mu_t)+C^{\rm SRR}_2(\mu_t)\right]
\end{eqnarray}
where~\cite{BUJAWE} $r=0.985$ and the factors $\bar P^a_i$ are given
in~\cite{BUCHROSL,BUJAUR}.

\begin{figure}[htbp]
\begin{center}
\begin{tabular}{p{0.48\linewidth}p{0.48\linewidth}}
\mbox{\epsfig{file=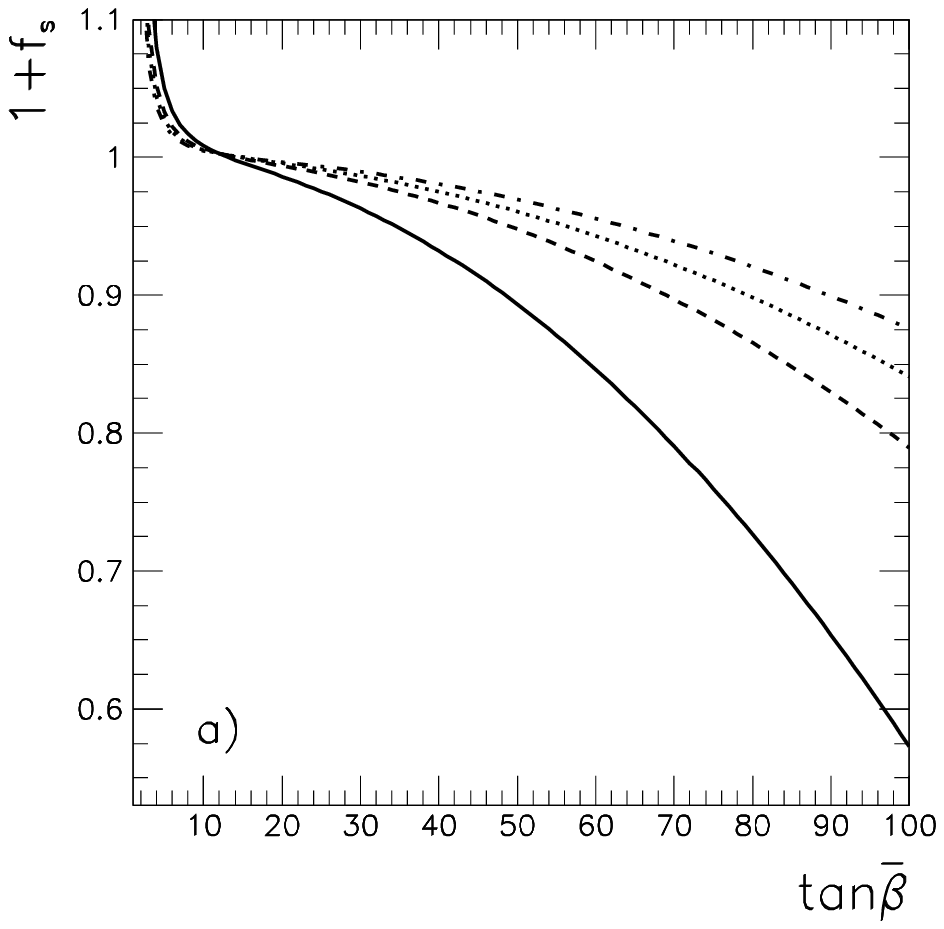,width=\linewidth}}&
\mbox{\epsfig{file=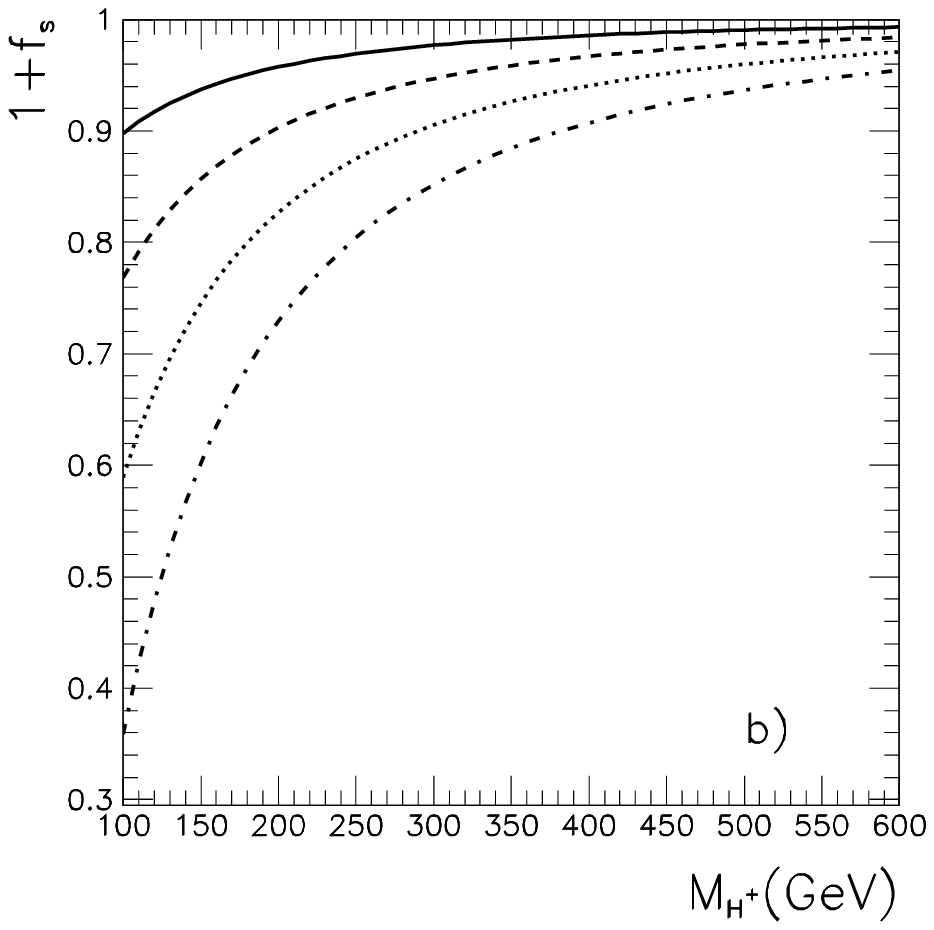,width=\linewidth}}\\
\end{tabular}
\vspace*{-5mm}
\caption{$1+f_s$ in the 2HDM(II): a) as a function of $\tan\beta$ 
  for $M_{H^+}=$ (from below) $150$, $250$, $300$ and $350$ GeV and b)
  as a function of $M_{H^+}$ for $\tan\beta=$ (from above) $40$,
  $60$, $80$ and $100$.
\label{fig:bcrs9}}
\end{center}
\end{figure}

Let us consider first the 2HDM(II). At 1-loop only the charged scalar
is relevant for the box diagrams contributing to \KK and \BB mixing.
For large $\tan\beta$ and $M_{H^+}\approx m_t$ the leading terms of
such contributions to the Wilson coefficients $C_i$ are of the order
of (see~\cite{BUCHROSL} for the complete expressions):
\begin{eqnarray}
\delta^{(+)}C^{\rm VLL}_1\sim{4\over3}\cot^2\beta ,\phantom{aa}
\delta^{(+)}C^{\rm LR}_2
\sim-{8\over3}{m_{d_I}m_{d_J}\over m^2_t}\tan^2\beta
\label{eqn:WHorder}
\end{eqnarray}
for diagrams with $W^\pm H^\mp$, and
\begin{eqnarray}
\delta^{(+)}C^{\rm VLL}_1\sim
{1\over3}{m^2_t\over M^2_W}\cot^2\beta  ,\phantom{aa}
\delta^{(+)}C^{\rm VRR}_1\sim{1\over3}{m^2_{d_I}m^2_{d_J}\over M^2_Wm^2_t}
\tan^4\beta  ,\phantom{aa}
\delta^{(+)}C^{\rm LR}_1\sim 0\nonumber\\
\delta^{(+)}C^{\rm SLL}_1\sim 0 ,\phantom{aa}
\delta^{(+)}C^{\rm SRR}_1\sim 0 ,\phantom{aa}
\delta^{(+)}C^{\rm LR}_2
\sim-{4\over3}{m_{d_I}m_{d_J}\over M^2_W}\tan^2\beta
\phantom{aaa}
\label{eqn:HHorder}
\end{eqnarray}
for diagrams with $H^\pm H^\mp$. For large $\tan\beta$ the biggest
contribution appears in $\delta^{(+)}C^{\rm LR}_2$ and is further
enhanced, compared to the SM amplitude, by the QCD renormalization
effects~\cite{BUJAWE}.  It is of the opposite sign than given by the
$tW^\pm$ box diagram and can be significant only for the $\bar
B^0_s$-$B^0_s$ mixing (so that it can compete with the SM contribution
- see fig.~\ref{fig:bcrs9}), for which it is of the order of
\begin{eqnarray}
\delta^{(+)}C^{\rm LR}_2\approx
-{2m_s(\mu_t)m_b(\mu_t)\over M^2_W}\tan^2\beta
\approx-0.14\times\left({\tan\beta\over50}\right)^2~.
\label{eqn:BBorder}
\end{eqnarray}
Similar contributions for $\bar B^0_d$-$B^0_d$ and \KK transitions are
suppressed by factors $m_d/m_s$ and $m_d/m_b$, respectively, and thus
very small.
Unfortunately, the new CLEO experimental result for the process
$BR(B\rightarrow X_s\gamma)=(3.03\pm0.40\pm0.26)\times10^{-4}$ set the
bound $M_{H^+}\simgt380$ GeV~\cite{GAMI}. Thus, in the 2HDM(II) for the
still allowed range of charged Higgs boson mass the effects in $\Delta
M_{B_s}$ cannot be large.

As a second realistic GMFV model we consider the MSSM with the CKM as
the only source of flavour/CP violation.  In the limit of heavy
sparticles (practically realized already for $M_{SUSY}\simgt500$ GeV)
the 1-loop box diagrams involving charginos and stops are negligible.
However, for large $\tan\beta$ even if sparticles are heavy they can
still compensate the $H^\pm$ contribution to the $b\rightarrow
s\gamma$ amplitude allowing for the existence of a light, $\sim{\cal
  O}(150$ GeV), charged Higgs boson~\cite{DEGAGI,CAGANIWA}.  From
fig.~\ref{fig:bcrs9} it follows therefore that, even for
$\tan\beta\simlt 50$ and already at the 1-loop level the contribution
of the MSSM Higgs sector to the $C_2^{\rm LR}$ Wilson coefficient can
be non-negligible.

At the 2-loop level in the MSSM one has to take into account the
dominant 2-loop electroweak corrections, modifying the Yukawa-type
interactions.  Even for heavy sparticles these corrections can
significantly modify~\cite{DEGAGI} the 2HDM(II) relations between the
masses $m_{d_I}$ and the Yukawa couplings $Y_d^{I}$,
as well as induce additional, $\propto\tan\beta$, terms in the charged
Higgs-quark couplings~\cite{CIDEGAGI,DEGAGI}.  For the $\Delta F=2$
processes the most important (for non-negligible mixing of the top
squarks) is however their third effect, that is the generation of the
flavour non-diagonal, $\tan\beta$ enhanced couplings of the neutral
Higgs bosons to down-type quarks~\cite{HAPOTO,BAKO}.  Additional
contributions to $C^{\rm LL}_1$, $C^{\rm RR}_1$ and $C^{\rm LR}_2$ are
then generated by the double penguin diagrams involving the neutral
Higgs bosons exchanges.  The dominant terms obtained from such
contributions are
\begin{eqnarray}
&&\delta^{(0)}C^{\rm LL}_1=-{\alpha_{EM}\over4\pi s^2_W}{m^4_t\over M^4_W}
m_{d_J}^2 X_{tC}^2 \tan^4\beta ~{\cal F}_-\phantom{aa}\nonumber\\ 
&&\delta^{(0)}C^{\rm RR}_1=-{\alpha_{EM}\over4\pi s^2_W}{m^4_t\over M^4_W}
m_{d_I}^2 X_{tC}^2 \tan^4\beta ~{\cal F}_-\phantom{aa}
\label{eqn:babucor}
\\ 
&&\delta^{(0)}C^{\rm LR}_2=-{\alpha_{EM}\over2\pi s^2_W}{m^4_t\over M^4_W}
m_{d_J}m_{d_I} X_{tC}^2 \tan^4\beta ~{\cal F}_+~.\phantom{aa}\nonumber
\end{eqnarray}
with $X_{tC}$ given by (the matrices $Z_+$ and $Z_-$ are defined
in~\cite{ROS}):
\begin{eqnarray}
X_{tC} = \sum_{j=1}^2Z_+^{2j}Z_-^{2j}{A_t\over m_{C_j}}
H_2\left({M^2_{\tilde t_1}\over m^2_{C_j}},
{M^2_{\tilde t_2}\over m^2_{C_j}}\right),
\end{eqnarray}
where the function $H_2(x,y)$ is given in~\cite{BUCHROSL} and the
factor ${\cal F}_\mp$ reads as
\begin{eqnarray}
{\cal F}_\mp\equiv\left[{\cos^2\alpha\over M^2_H} 
+ {\sin^2\alpha\over M^2_h} \mp {\sin^2\beta\over M^2_A}\right]
\end{eqnarray}
$\tan^4\beta$ factor in eq.~(\ref{eqn:babucor}) appears because the
dominant part of the effective flavour off-diagonal Yukawa coupling is
given~\cite{CHSL} by the flavour-changing chargino contribution to the
$d$-quark self energy ($\propto \tan\beta$), multiplied by the
tree-level Yukawa coupling (also $\propto \tan\beta$).  Two powers of
the external light quark masses in~(\ref{eqn:babucor}) are canceled
by the fermion propagators on internal lines.

\begin{figure}[b!]
\begin{center}
\vspace*{-8mm}
  \epsfig{file=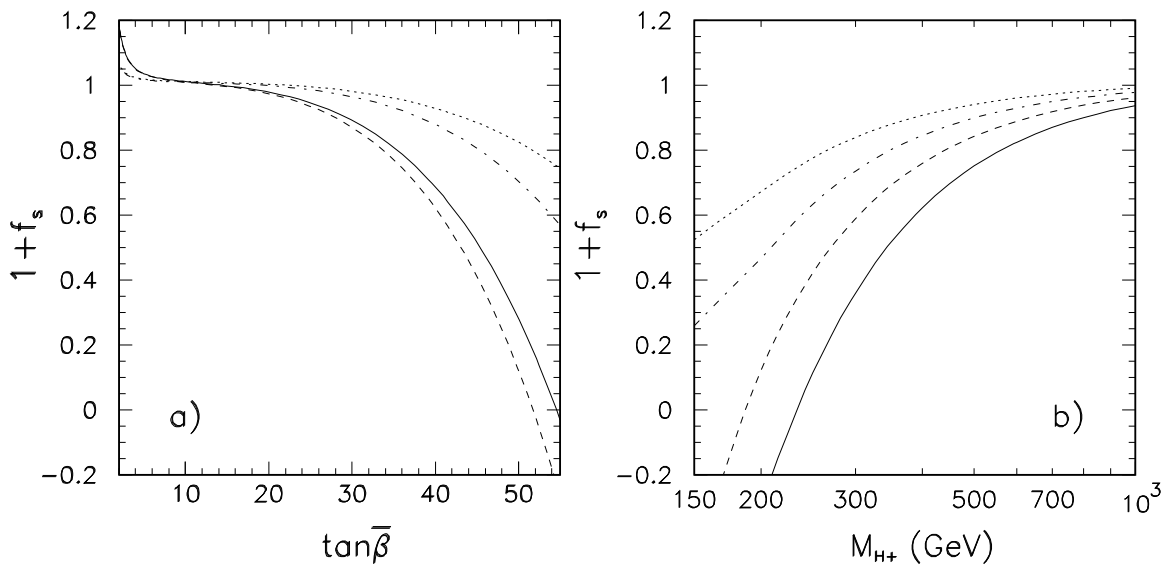,width=\linewidth}
\end{center}
\vspace*{-6mm}
\caption{\protect $1+f_s$ in the MSSM for lighter chargino mass
  750 GeV, $M_2/\mu=-0.5$ and stop masses (in GeV) (500,850),
  (700,1000), (500,850) and (600,1100) (solid, dashed, dotted and
  dot-dashed lines, respectively) as a function of a) $\tan\beta$ and
  b) $M_{H^+}$.  In panel a) solid and dashed (dotted and dot-dashed)
  lines correspond to $M_{H^+}=200$ $(600)$ GeV, and in panel a) solid
  and dashed (dotted and dot-dashed) lines correspond to
  $\tan\beta=50$ $(35)$.}
\label{fig:bcrs13}
\end{figure}

For $\tan\beta\gg 1$, ${\cal F}_-$ is close to zero~\cite{BAKO}, but
the correction $\delta^{(0)}C^{\rm LR}_2$ is proportional to ${\cal
  F}_+$ which is not suppressed in this limit.  Approximating the
dimensionless factor $X_{tC}$ by unity, it is easy to see that in the
case of the \BBs mixing this correction can be for $\tan\beta\sim 50$
and $M_{H^+}\sim 200$ GeV as large as $\delta^{(0)}C^{\rm
  LR}_2\sim2.5$ i.e. of the same order of magnitude as the SM
contribution to $C^{\rm VLL}_1$.  This is illustrated in
fig.~\ref{fig:bcrs13}.

An important features of the double penguin contribution are: {\sl i)}
its fixed negative sign, the same as the sign of the dominant effects
of the charged Higgs box diagrams at large $\tan\beta$; {\sl ii)} its
strong dependence on the left-right mixing of the top squarks - from
fig.~\ref{fig:bcrs13} it follows that large values of the stop mixing
parameter $A_t$ are excluded already by the present experimental data;
{\sl iii)} it does not vanish when the mass scale of the sparticles is
increased uniformly - large effects decreasing $\Delta M_{B_s}$ can be
present also for the heavy sparticles provided the Higgs boson masses
remain low.\\[1mm]
Summarizing, we found that:\\[1mm]
{$\bullet$} The largest effects of new contributions for large
$\tan\beta$ are seen in $\Delta M_{B_s}$. The corresponding
contributions to $\Delta M_{B_d}$ and $\varepsilon$ are strongly
suppressed by the smallness of $d$-quark mass.\\[1mm]
{$\bullet$} The dominant contributions to $\Delta M_{B_s}$ for large
$\tan\beta$ originate from double penguin diagrams involving neutral
Higgs particles and, to a lesser extent, in the box diagrams
with charged Higgs exchanges.\\[1mm]
{\bf $\bullet$} The contribution of double penguins grows like
$\tan^4\beta$ and interferes destructively with the SM contribution,
suppressing considerably $\Delta M_{B_s}$. It depends strongly on stop
mixing, excluding large values of the mixing parameter $A_t$.

\section*{Acknowledgments}

This work was supported in part by the German Bundesministerium f\"ur
Bildung and Forschung under the contract 05HT1WOA3, by the Polish
State Committee for Scientific Research grant 2 P03B 060 18 for years
2000-2001 and by the EC Contract HPRN-CT-2000-00148 for years
2000-2004.

\end{document}